\documentclass[english,preprint]{aastex}
\usepackage[T1]{fontenc}
\usepackage[latin9]{inputenc}
\usepackage{amsbsy}
\usepackage{amssymb}
\usepackage{esint}
\usepackage{babel}
\begin{document}

\title{Non-Zeeman Circular Polarization of Molecular Maser Spectral Lines}

\author{Martin Houde}

\affil{Department of Physics and Astronomy, The University of Western Ontario,
London, ON, N6A 3K7, Canada}

\affil{Division of Physics, Mathematics and Astronomy, California Institute
of Technology, Pasadena, CA 91125 }
\begin{abstract}
We apply the anisotropic resonant scattering model developed to explain
the presence of non-Zeeman circular polarization signals recently
detected in the $^{12}\mathrm{CO}\;\left(J=2\rightarrow1\right)$
and $\left(J=1\rightarrow0\right)$ transitions in molecular clouds
to Stokes $V$ spectra of SiO $v=1$ and $v=2$, $\left(J=1\rightarrow0\right)$
masers commonly observed in evolved stars. It is found that the observed
antisymmetric ``S'' and symmetric ``$\cup$'' or ``$\cap$''
shaped spectral profiles naturally arise when the maser radiation
scatters off populations of foreground molecules located outside the
velocity range covered by the background maser radiation. Using typical
values for the relevant physical parameters, it is estimated that
magnetic field strengths on the order of a few times 15 mG are sufficient
to explain the observational results found in the literature.
\end{abstract}


\section{Introduction\label{sec:Introduction}}

The recent discovery of circular polarization in $^{12}\mathrm{CO}$
spectral lines in molecular clouds \citep{Houde2013,Hezareh2013}
have opened up the possibility of developing a new method for mapping
magnetic fields. Until now the measurement of circular polarization
in spectral line profiles in the denser parts of molecular clouds
and star-forming regions have been limited to the few Zeeman sensitive
molecules that have strong enough line intensities to yield a measurable
broadening in line profiles (e.g., OH and CN; see \citealt{Guesten1994,Crutcher1999,Falgarone2008}).
It follows that further polarization measurements on strong maser
lines from several molecular species have also been particularly useful
for studies of this kind (e.g., $\mathrm{H_{2}O}$, $\mathrm{CH_{3}OH}$
class I and II, $\mathrm{OH}$, and SiO;\textbf{ }see \citealt{Fish2007,Watson2009,Vlemmings2012,Sarma2012}
for recent reviews). Zeeman measurements result in direct estimates
of magnetic field strengths (usually the line of sight component).
However, the very low levels of polarization thus detected in the
corresponding Stokes $V$ spectra imply that Zeeman studies can, so
far, only be realized on a relatively small number of sources. It
also follows that despite the fact that all other techniques used
to study magnetic fields in weakly ionized media, either based on
polarimetry (e.g., \citealt{CF1953,Hildebrand2000,Hildebrand2009,Heyer2008,Houde2011,Houde2009,Houde2004})
or not \citep{Houde2000a,Houde2000b,Houde2001,Li2008}, only give
indirect information about magnetic fields, they have nonetheless
played an essential role for such studies.

The $^{12}\mathrm{CO}\;\left(J=2\rightarrow1\right)$ Stokes $V$
spectrum measured by \citet{Houde2013} at the peak intensity position
in Orion KL, with the Caltech Submillimeter Observatory \citep{Hezareh2010},
showed a clear detection of polarization at levels on the order $1\%$
away from the line core. Although their proposed anisotropic resonant
scattering process could  account for these levels of polarization,
there remained two issues that could not be satisfactorily assessed
at the time. First, within the resonant scattering model the presence
of circular polarization in spectral lines arises from the conversion
of linearly polarized radiation emanating from a population of molecules
in the background. This polarization transformation occurs when this
incident radiation scatters off a second population of the same molecular
species in the foreground; a process that could not be experimentally
verified. Second, it was also found that the Stokes $V$ line shapes
predicted by the resonant scattering model tend to display an intrinsic
antisymmetric profile, as do the Zeeman effect and other scattering
processes (e.g., \citealt{Deguchi1985}), while the observations of
$^{12}\mathrm{CO}\;\left(J=2\rightarrow1\right)$ in Orion KL did
not.

The recent work of \citet{Hezareh2013} on SNR IC 443 (G) have, however,
provided strong evidence that the circular polarization signals they
detected in $^{12}\mathrm{CO}\;\left(J=2\rightarrow1\right)$ and
$\left(J=1\rightarrow0\right)$ over the whole source (spanning between
$1\arcmin$ to $2\arcmin$ in size) indeed result from a conversion
of linear to circular polarization. These observations were obtained
with the IRAM 30-m telescope, where circular polarization was detected
at levels of up to approximately $1\%$, while linear polarization
was generally weaker at a few tenths of a percent. In their analysis,
\cite{Hezareh2013} compared the polarization angles (PA) of the $^{12}\mathrm{CO}$
linear polarization with those from a map of dust emission at 345
GHz obtained with PolKa at APEX. The observed difference in PA between
the two sets of data was perfectly consistent with what would be expected
if the background $^{12}\mathrm{CO}$ linear polarization (aligned
with the dust polarization) was partially converted to circular polarization
through the anisotropic resonant scattering process of \cite{Houde2013}.
That is, as this polarization conversion process takes place, the
scattered and remaining $^{12}\mathrm{CO}$ linear polarization is
rotated away from the PA of the incident polarization (and therefore
that of the dust) in a predictable manner. Accordingly, it was then
found that although the alignment between co-located linear polarization
vectors in the molecular and dust polarization maps was almost inexistent,
the reinsertion of the measured circular polarization in the linear
polarization signals (i.e., by reversing the conversion of linear
to circular polarization) revealed a very strong agreement between
the corrected $^{12}\mathrm{CO}$ linear polarization vectors and
those from the dust continuum. Still, it remained that the Stokes
$V$ spectra measured in SNR IC 443 (G) did not reveal clear antisymmetric
line profiles. 

As explained in \citet{Houde2013} the shape of a given Stokes $V$
spectrum produced by the anisotropic resonant scattering process can
be very difficult to predict for turbulent molecular lines, such as
those measured in Orion KL and SNR IC 443 (G). This is because there
is not one well-defined propagation path for the detected radiation.
That is, the nature of resonant scattering itself makes it likely
that this radiation can emanate from several directions ``behind''
the scattering molecules before scattering into the telescope beam.
It is also to be expected that background photons emanating from different
paths will experience differing conditions that will affect their
scattered amplitude and conversion to circular polarization (e.g.,
changes in the angle between the incident linear polarization and
the magnetic field aligning the scattering molecules). It follows
from these considerations that masers, in view of their nature and
simpler propagation characteristics, may provide a particularly interesting
test for Stokes $V$ line shapes predicted by the anisotropic resonant
scattering model. 

This is what we endeavor to accomplish in this paper, where we apply
the model of \citet{Houde2013} to the problem of maser radiation
and compare its predictions to observational results existing in the
literature. More precisely, we aim to investigate whether the line
shapes and intensities of Stokes $V$ spectra from SiO $v=1$ and
$v=2$, $\left(J=1\rightarrow0\right)$ masers (at 43.1 GHz and 42.8
GHz, respectively) detected in the AGB star IK Tau by \citet{Cotton2011}
can be readily reproduced by the model of \citet{Houde2013}. In what
follows, we present a brief review of the anisotropic resonant scattering
process in Section \ref{sec:ARS}, show results from numerical calculations
tailored to the SiO 42.8 GHz, $v=2$, $\left(J=1\rightarrow0\right)$
maser line in Section \ref{sub:SiO}, and discuss their comparison
with the observations of \citet{Cotton2011} in Section \ref{sec:Discussion}.
Finally, we end with a short summary in Section \ref{sec:Conclusion}.

\section{Linear to Circular Polarization Conversion through Anisotropic Resonant
Scattering\label{sec:ARS}}

Following \citet{Houde2013}, we introduce an incident background
radiation state $\left|\psi_{0}\right\rangle $ linearly polarized
at an angle $\theta$ with the plane-of-the-sky component of the foreground
magnetic field 

\begin{equation}
\left|\psi_{0}\right\rangle =\alpha_{0}\left|n_{\Vert}\right\rangle +\beta_{0}\left|n_{\bot}\right\rangle ,\label{eq:psi0}
\end{equation}

\noindent where $\alpha_{0}=\cos\left(\theta\right)$ and $\beta_{0}=\sin\left(\theta\right)$.
The $n$-photon states $\left|n_{\Vert}\right\rangle $ and $\left|n_{\bot}\right\rangle $
are orthonormal and polarized in directions parallel and perpendicular
to the foreground magnetic field, respectively. We assume that the
incident radiation characterized by equation (\ref{eq:psi0}) emanates
from a population of a given molecular species in the background medium
while another population of the same molecule responsible for the
scattering of this incident radiation coexists with, and is spatially
aligned by, the foreground magnetic field. As was shown in \citet{Houde2013},
after undergoing a large number of anisotropic resonant scattering
events with the foreground population the incident radiation is transformed
to a new state 

\begin{equation}
\left|\psi\right\rangle \simeq\alpha_{0}e^{i\phi}\left|n_{\Vert}\right\rangle +\beta_{0}\left|n_{\bot}\right\rangle ,\label{eq:psi}
\end{equation}

\noindent where $\phi$ is the relative phase shift induced between
the two linear polarization states in the process. This final state
of the scattered radiation is what we would seek to measure with a
telescope. Accordingly, it is straightforward to show that the corresponding
normalized Stokes parameters defining the complete polarization state
of the scattered signal are%
\footnote{We use the IAU convention for circular polarization, i.e., Stokes
$v=\mathrm{RHC}-\mathrm{LHC}$, with RHC the right-handed circular
polarization intensity having its electric vector rotating counter-clockwise
as seen by the observer, etc. \citet{Houde2013} used the IEEE convention,
which defines Stokes $v$ as the negative of the above. %
}

\begin{eqnarray}
q & = & \alpha_{0}^{2}-\beta_{0}^{2}\label{eq:q}\\
u & = & 2\alpha_{0}\beta_{0}\cos\left(\phi\right)\label{eq:u}\\
v & = & 2\alpha_{0}\beta_{0}\sin\left(\phi\right).\label{eq:v}
\end{eqnarray}

\noindent For these equations we chose a system of reference aligned
with axes parallel and perpendicular to the foreground magnetic field.
More precisely, the linear polarization angle

\begin{equation}
\chi=\frac{1}{2}\arctan\left(\frac{u}{q}\right)\label{eq:chi}
\end{equation}

\noindent is measured from the axis defined by the orientation of
the magnetic field. The Stokes parameters $q_{0}$, $u_{0}$, and
$v_{0}$($=0$) of the incident background radiation (i.e., pre-scattering),
as well as the incident linear polarization angle $\chi_{0}$, are
obtained by setting $\phi=0$ in equations (\ref{eq:q})-(\ref{eq:v}).
We then find that, in this particular reference frame, 

\begin{eqnarray}
q & = & q_{0}\label{eq:q-q0}\\
u & = & u_{0}\cos\left(\phi\right)\label{eq:u-u0}\\
v & = & u_{0}\sin\left(\phi\right).\label{eq:v-u0}
\end{eqnarray}

\noindent It thus becomes clear from these relations that as a relative
phase shift $\phi$ builds up linear polarization is transferred from
the incident Stokes $u_{0}$ parameter to the scattered circular polarization
Stokes $v$. The total amount of polarization is conserved in the
process. It is important to note that the linear polarization angle
is also transformed with

\begin{equation}
\tan\left(2\chi\right)=\cos\left(\phi\right)\tan\left(2\chi_{0}\right),\label{eq:chi-chi0}
\end{equation}

\noindent i.e., as linear polarization is converted to circular polarization
the linear polarization angle rotates away from its original orientation. 

This is the effect that was observed by \citet{Hezareh2013} for the
$^{12}\mathrm{CO}\;\left(J=2\rightarrow1\right)$ and $\left(J=1\rightarrow0\right)$
transitions in SNR IC 443 (G), as previously mentioned. They were
able to recover $\chi_{0}$ at every position on their maps by successively
testing out different orientations for the foreground magnetic field.
This allowed them to: \emph{i)} single out new reference frames to
calculate the Stokes parameters $q$($=q_{0}$) and $u$ through a
rotation of their measured parameters (in the equatorial coordinate
system, Stokes $v$ is independent of the reference frame), \emph{ii)}
calculate the relative phase shift with $\phi=\arctan\left(v/u\right)$
(see equations (\ref{eq:u}) and (\ref{eq:v})), \emph{iii)} determine
the incident Stokes parameter

\begin{equation}
u_{0}=u\cos\left(\phi\right)+v\sin\left(\phi\right),\label{eq:recover-u0}
\end{equation}

\noindent and finally \emph{iv)} calculate $\chi_{0}$ with a relation
similar to equation (\ref{eq:chi}) using $q_{0}$ and $u_{0}$. The
correct orientation for the foreground magnetic field was readily
identified when the new angles $\chi_{0}$ aligned to a very high
degree with the PA obtained with their complementary dust polarization
measurements. \citet{Hezareh2013} were thus able to confirm that
the presence of circular polarization in their $^{12}\mathrm{CO}$
spectral lines indeed originates from a conversion of incident background
linear polarization.

The strength of the polarization conversion effect depends on the
physical conditions found in both the background region where the
incident linear polarized radiation emanates from and the volume occupied
by the foreground scattering gas. It is therefore important to distinguish
between the two corresponding sets of parameters entering in the different
equations that follow. We also emphasize that the conditions found
in the two regions (i.e., background and foreground) can be significantly
different; this is certainly the case for the excitation temperatures
and gas densities. This is especially important to keep in mind for
the case treated in this paper since the incident radiation we will
consider results from masers located in the background, implying,
for example, the presence of a population inversion. The foreground
molecular population is in no such way constrained. In fact, although
unlikely to be perfectly realized, we will assume that Local Thermodynamic
Equilibrium (LTE) applies for the determination of level populations
in the foreground scattering gas.

According to the anisotropic resonant scattering model of \citet{Houde2013}
the relative phase shift can be evaluated as a function of the frequency
$\omega$ of the incident radiation with

\begin{equation}
\phi\left(\omega\right)\simeq\omega_{Z}^{2}\sin^{2}\left(\iota\right)\tau\mathcal{V_{\mathrm{int}}}\frac{ng_{\mathrm{l}}e^{-E_{\mathrm{l}}/kT_{\mathrm{ex}}}}{Q\left(T_{\mathrm{ex}}\right)}\frac{3\pi c^{3}A_{\mathrm{ul}}}{4\hbar\omega_{0}^{3}\omega^{2}}\sqrt{u\left(\omega\right)u^{\prime}\left(\omega\right)}\, I\left(\omega\right),\label{eq:phi}
\end{equation}

\noindent where

\begin{eqnarray}
I\left(\omega\right) & = & \int\left\{ x^{2}\left(x-\omega\right)\left[3\left(x-\omega\right)^{2}-\gamma_{\mathrm{ul}}^{2}-\omega_{Z}^{2}\right]/\left[\left(x-\omega\right)^{2}+\gamma_{\mathrm{ul}}^{2}\right]\right.\nonumber \\
 &  & \left.+\left(x-\omega\right)\left(\omega^{2}-3x^{2}\right)+\gamma_{\mathrm{ul}}^{2}\left(3x-\omega\right)+\omega_{Z}^{2}\left(x+\omega\right)\rule{0in}{2.5ex}\right\} \frac{h\left(x\right)}{\Delta}dx,\label{eq:I(w)}
\end{eqnarray}

\noindent with $\Delta=\left[\left(x+\omega_{Z}-\omega\right)^{2}+\gamma_{\mathrm{ul}}^{2}\right]\left[\left(x-\omega_{Z}-\omega\right)^{2}+\gamma_{\mathrm{ul}}^{2}\right]$.
In equations (\ref{eq:phi}) and (\ref{eq:I(w)}) $\omega_{0}$ is
the frequency of the transition between the upper (u) and lower (l)
energy levels in the rest frame of the scattering molecules (i.e.,
the frequency of the $\pi$-transition), $\omega_{Z}$ the Zeeman
splitting due to the foreground magnetic field, $\gamma_{\mathrm{ul}}$
the total relaxation rate pertaining to the upper and lower states
for the scattering molecules, $n$ their total volume density, $A_{\mathrm{ul}}$
the Einstein coefficient for the transition, $u\left(\omega\right)$
and $u^{\prime}\left(\omega\right)$ the energy densities of the incident
and scattered radiation ($u\left(\omega\right)$ corresponds to Stokes
$u_{0}$ in the previous discussion), $h\left(x\right)$ the normalized
spectral profile of the population of scattering molecules, and $\iota$
the inclination angle of the foreground\textbf{ }magnetic field to
the line of sight. The population of the lower state for the scattering
molecules is calculated using the\textbf{ }degeneracy $g_{\mathrm{l}}$
and energy level $E_{\mathrm{l}}$, while LTE at an excitation temperature
$T_{\mathrm{ex}}$, at which the partition function $Q$ is evaluated,
was assumed for the corresponding term in equation (\ref{eq:phi}).
The volume of the region of interaction $\mathcal{V_{\mathrm{int}}}$,
where resonant scattering occurs, is constrained in directions perpendicular
to the radiation's direction of propagation by the lifetimes of the
two maser\textbf{ }states through $l_{\gamma}\approx c/\gamma_{\mathrm{ul}}^{\prime}$,
with $\gamma_{\mathrm{ul}}^{\prime}$ the inverse of the total relaxation
rate corresponding to the masing molecules. Along the direction of
propagation $\mathcal{V_{\mathrm{int}}}$ is limited to the smallest
of the physical size $l_{\mathrm{s}}$ of the region harboring the
scattering molecules and the effective mean path $l_{\mathrm{p}}$
of a photon. The time of interaction $\tau$ between the radiation
and the molecules is set by $1/\gamma_{\mathrm{ul}}^{\prime}$.

For the numerical calculations to be presented in the next section
we use the following relations for the total relaxation rate corresponding
to the upper and lower states of the maser transition and the effective
mean path of a photon

\begin{eqnarray}
\gamma_{\mathrm{ul}}^{\prime} & = & \gamma_{\mathrm{coll}}+\frac{\gamma_{\mathrm{rad}}}{2}\label{eq:gamma_ul}\\
l_{\mathrm{p}} & = & \left[\alpha_{\omega}\left(\alpha_{\omega}+\sigma_{\omega}\right)\right]^{-1/2},\label{eq:l_p}
\end{eqnarray}

\noindent where the collision rate of a maser molecule with the (dominant)
hydrogen molecules of density $n_{\mathrm{H}_{2}}$ is $\gamma_{\mathrm{coll}}=n_{\mathrm{H}_{2}}\left\langle \sigma v\right\rangle $,
with the momentum-rate transfer coefficient $\left\langle \sigma v\right\rangle $
evaluated at the kinetic temperature of the gas harboring the maser
$T_{\mathrm{kin}}$ \citep{Shull1990}. The total radiative decay
rate $\gamma_{\mathrm{rad}}$ includes all the related rates for the
lower and upper states of the background maser population\textbf{
}taken separately, as well as the emission (spontaneous and stimulated)
and absorption rates for radiative transitions between them \citep{Letokhov2009}.
We have a similar equation for $\gamma_{\mathrm{ul}}$ the relaxation
rate of the scattering molecules, but using the corresponding hydrogen
molecules density, excitation temperature $T_{\mathrm{ex}}$, and
radiative rates at that location. The coefficients of absorption \citep{Rybicki1979}
and resonant scattering \citep{Grynberg2010} integrated over the
foreground\textbf{ }molecular population of the lower state are respectively

\begin{eqnarray}
\alpha_{\omega} & = & \frac{ng_{\mathrm{u}}e^{-E_{\mathrm{l}}/kT_{\mathrm{ex}}}}{Q\left(T_{\mathrm{ex}}\right)}\frac{\pi^{2}c^{2}A_{\mathrm{ul}}}{\omega^{2}}\frac{\hbar\omega}{kT_{\mathrm{ex}}}h\left(\omega\right)\label{eq:alpha_w}\\
\sigma_{\omega} & = & \frac{ng_{\mathrm{l}}e^{-E_{\mathrm{l}}/kT_{\mathrm{ex}}}}{Q\left(T_{\mathrm{ex}}\right)}\frac{3c^{2}}{\omega^{2}}4\pi^{3}\gamma_{\mathrm{ul}}^{\prime}h\left(\omega\right).\label{eq:sigma_w}
\end{eqnarray}

We note, finally, that the term $\omega_{Z}^{2}\sin^{2}\left(\iota\right)$
in equation (\ref{eq:phi}) makes clear the explicit dependency of
the polarization conversion effect on the square of the strength of
the plane-of-the-sky component of the foreground magnetic field. This
will therefore be another parameter entering in the numerical calculations
to follow.

\section{Circular Polarization Spectral Line Profiles of SiO $\boldsymbol{v=2}$,
$\boldsymbol{\left(J=1\rightarrow0\right)}$ Masers\label{sub:SiO}}

In a recent study of SiO $v=1$ and $v=2$, $\left(J=1\rightarrow0\right)$
masers (at 43.1 GHz and 42.8 GHz, respectively) in the AGB star IK
Tau \citet{Cotton2011} presented, among other results, several Stokes
$V$ spectra obtained at high spatial resolution with the Very Long
Baseline Array. Their results revealed high levels of circular polarization
often amounting to a significant fraction of the strong linear polarization
detected at corresponding positions. It is expected that the large
levels of linear polarization detected for these lines are probably
a consequence of anisotropic pumping, as they would otherwise require
an implausible level of maser saturation \citep{Watson2009}. Although
some of the Stokes $V$ spectral lines reported by \citet{Cotton2011} 
displayed typical Zeeman-like antisymmetric ``S'' shaped profiles,
several symmetric ``$\cup$'' or ``$\cap$'' shaped profiles were
also observed. Considering these circular polarization line profiles
and levels \citet{Cotton2011} concluded that their observations could
not be explained through the Zeeman effect and related models \citep{Elitzur1996}.
They further considered the alternative scenario of \citet{Wiebe1998}
for the conversion of linear to circular polarization based on population
imbalance \citep{Deguchi1985}, but were also unable to successfully
account for their observations with this model.

In this section we endeavor to test whether the anisotropic resonant
scattering model of \citet{Houde2013} can reproduce the observations
of \citet{Cotton2011}. To do so implies establishing a mechanism
through which symmetric ``$\cup$'' and ``$\cap$'' shaped Stokes
$V$ profiles can be produced, in addition to the antisymmetric ``S''
shaped. This problem was not resolved in \citet{Houde2013}. 

To that end, it is first instructive to explain how the antisymmetric
profiles arises in the resonant scattering model. For this we consider
the basic resonance integral that arises from the interaction between
an incident photon of frequency $\omega$ with scattering molecules
for a given transition (e.g., the $\pi$-transition; see equations
(34) or (40) of \citealt{Houde2013})

\begin{equation}
J\left(\omega\right)\simeq\int\frac{x-\omega}{\left(x-\omega\right)^{2}+\gamma_{\mathrm{ul}}^{2}}h\left(x\right)dx.\label{eq:J(w)}
\end{equation}

\noindent We note from this equation that for a symmetric profile
$h\left(x\right)$ the scattering amplitude will have differing signs
for frequencies on opposite sides of its center. Accordingly, the
amplitude becomes zero when $\omega$ is located right at the center
of $h\left(x\right)$, as there is then a net cancellation of scattering
from molecules residing at lower and higher frequencies (or velocities).
The scattering amplitude will be non-zero for any other photon frequencies,
being positive or negative depending whether $\omega$ is on the low
or high side of $h\left(x\right)$, respectively. This behavior is
responsible for the appearance of an antisymmetric ``S'' shaped
line profile when the spectral distribution of the incident radiation
($u\left(\omega\right)$ in equation (\ref{eq:phi})) is also symmetric
about the center of $h\left(x\right)$.

We should also be aware, however, that the outcome would be completely
different if the profiles of the incident radiation $u\left(\omega\right)$
and the scattering molecules population $h\left(x\right)$ overlap
minimally in velocity space. Indeed, a ``$\cup$'' or ``$\cap$''
shaped scattered profile would result if $h\left(x\right)$ was entirely
located on only one side of the incident radiation profile with little
to no overlap. For the case of a maser, this implies a resonant scattering
that is external to the masing process. That is, while molecules in
the foreground residing at velocities covered by the maser could simply
partake in the masing process, other molecules outside that velocity
range would act as resonant scatterers. The remaining question is
whether the scattering would be strong enough to efficiently convert
incident linear polarization to circular polarization.

We use the model of \citet{Houde2013} summarized in equations (\ref{eq:phi})
to (\ref{eq:sigma_w}) above to determine whether it can produce Stokes
$V$ spectra consistent with the observations of \citet{Cotton2011}.
For this we have used the model developed by \citet{Decin2010} for
the circumstellar envelope (CSE) of IK Tau. The following parameters
were used for the background gas where the maser radiation originates
(in the inner CSE): $n_{\mathrm{H}_{2}}=10^{10}\;\mathrm{cm}^{-3}$,
$T_{\mathrm{kin}}=1000\;\mathrm{K}$; while in the foreground where
the resonant scattering takes place (5 AU further in the CSE): $n_{\mathrm{H}_{2}}=7.5\times10^{8}\;\mathrm{cm}^{-3}$,
$T_{\mathrm{ex}}=700\;\mathrm{K}$,\textbf{ }and $B=15\;\mathrm{mG}$.
In both regions the SiO abundance relative to $\mathrm{H}_{2}$ was
set to\textbf{ $1.6\times10^{-5}$}. We then applied these to the
SiO $v=2$, $\left(J=1\rightarrow0\right)$ 42.8 GHz maser transition
for which $g_{\mathrm{l}}=1$, $g_{\mathrm{u}}=3$, $E_{\mathrm{l}}/k=3527$
K, $Q\left(T_{\mathrm{ex}}\right)=1163$, and $A_{\mathrm{ul}}=3\times10^{-6}\:\mathrm{s}^{-1}$\citep{Muller2001}.
The Landé factor (relative to the nuclear magneton) is $g_{\mathrm{SiO}}\simeq-0.154$
to $-0.155$ for $v=0$, 1, and 2 \citep{Davis1974}. The incident
linear polarization radiation $u\left(\omega\right)$ was given a
Gaussian profile centered at $0\;\mathrm{km\, s}^{-1}$, with a FWHM
of $0.75\;\mathrm{km\, s}^{-1}$ and peak intensity of 1 Jy, approximately
matching the profiles presented in \citet{Cotton2011} (see their
Figure 5). Likewise, we also used a Gaussian profile $h\left(x\right)$
for two populations of scattering molecules of $1.5\;\mathrm{km\: s^{-1}}$
width (FWHM), centered at velocities (or frequencies) located at $\pm3\;\mathrm{km\: s^{-1}}$.
These distributions for the scattering molecules approximately match
the gas velocity profile at that position in the CSE, as determined
in \citet{Decin2010}. Depending on the cases studied in Figure \ref{fig:SiO_cp}
each population component can have a density $n=0$, 0.5, or 1 times
$1.2\times10^{3}\:\mathrm{cm}^{-3}$. We use a value $5\;\mathrm{s}^{-1}$
for the radiative decay rates of the (vibrationally excited) lower
and upper states and for the maser radiation we\textbf{ }assume that
the rate of stimulated emission is approximately equal to these decay
rates \citep{Wiebe1998,Watson2009}. Given these figures, and assuming
a similar rate of absorption transitions from the lower to the upper
state, we find that the collision rates are not significant in the
calculations of $\gamma_{\mathrm{ul}}$ and\textbf{ $\gamma_{\mathrm{ul}}^{\prime}$,}
and even more so the spontaneous emission rate $A_{\mathrm{ul}}$.
We thus calculate that $l_{\gamma}\approx c/\gamma_{\mathrm{ul}}^{\prime}\approx2.3\times10^{9}\;\mathrm{cm}$.
The absence of significant overlap between $u\left(\omega\right)$
and $h\left(x\right)$ (see equations (\ref{eq:l_p})-(\ref{eq:sigma_w}))
implies that the volume of the region of interaction cannot be constrained
along the direction of propagation by the mean path $l_{\mathrm{p}}$,
as it then becomes exceedingly large. Lacking information on the physical
size of the scattering region we adopt the value $l_{\mathrm{s}}=5\;\mathrm{AU}$
($7.5\times10^{13}\;\mathrm{cm}$). This does not appear to be an
unreasonable or excessive size since it is known that the SiO maser
ring (located inside the inner dust condensation zone) observed by
\citet{Cotton2011} has a diameter of approximately 8 AU (i.e., $0\farcs03$
at a distance of 265 pc), while the $11\;\micron$ interferometry
observations of \citet{Hale1997} revealed a succession of more distant
circumstellar shells surrounding IK Tau at radii of approximately
25 to 150 AU (i.e., $\sim0\farcs1$ to $0\farcs57$ at the same distance).

The results of our calculations are shown in Figure \ref{fig:SiO_cp},
where the circular polarization spectra of the $\mathrm{SiO}\;\left(v=2,\: J=1\rightarrow0\right)$
transition at 42.8 GHz is shown for five different scattering populations.
In all cases the left panel shows the relative phase shift $\phi_{\mathrm{SiO}}$
calculated with equation (\ref{eq:phi}) (solid curve; using the vertical
scale on the left), the incident linear polarization energy density
($u\left(v\right)$, broken curve; using the vertical scale on the
right), and the populations of scattering molecules ($h\left(v\right)$,
dotted-broken and hashed curves; normalized for display purposes).
The right panels show the resulting Stokes $V$ profile (solid curve)
and, once more, the incident linear polarization energy density ($u\left(v\right)$,
broken curve); both using the vertical scale on the right. All quantities
are plotted as a function of the velocity $v$. We also set the magnetic
field in the plane of the sky (i.e., $\iota=\pi/2$), $2\alpha_{0}\beta_{0}=1$
(i.e., $\theta=\pi/4$, resulting in $q_{0}=0$ and $u_{0}=1$), and
$u\left(v\right)\simeq u^{\prime}\left(v\right)$ for simplicity in
computing $\phi_{\mathrm{SiO}}$. The computed Stokes $V$ spectra
are therefore of the form

\begin{equation}
V\left(v\right)\propto e^{-\frac{1}{2}\left(\frac{v}{\Delta v}\right)^{2}}\sin\left[\phi_{\mathrm{SiO}}\left(v\right)\right],\label{eq:Stokes-V}
\end{equation}

\noindent where the Gaussian profile is associated with $u\left(v\right)$. 

The top panel of Figure \ref{fig:SiO_cp} shows a clear example of
a symmetric ``$\cup$'' shaped Stokes $V$ profile resulting from
a minimal overlap in velocity space between the population of scattering
molecules and the incident radiation. In this case the foreground
SiO molecules (hashed profile) are centered at $-3\;\mathrm{km\: s^{-1}}$
only and have a total density of $1.2\times10^{3}\:\mathrm{cm}^{-3}$.
As can be seen, even a relatively weak magnetic field of 15 mG is
sufficient to bring a significant relative phase shift of approximately
$-1.5\;\mathrm{rad}$ at $\simeq-0.2\;\mathrm{km\, s}^{-1}$. The
resulting Stokes $V$ intensity profile (right panel) is almost as
strong as the incident linear polarization signal. In the second panel
from the top an additional scattering population is added at $3\;\mathrm{km\: s^{-1}}$
but at half the density (i.e., $0.6\times10^{3}\:\mathrm{cm}^{-3}$).
This foreground population provides scattering of opposite sign and
is therefore responsible for the appearance of a ``positive'' shoulder
on the red side of the $\phi_{\mathrm{SiO}}$ and Stokes $V$ profiles.
The middle panel has equal foreground molecular populations on either
sides of the incident radiation spectrum resulting in antisymmetric
``S'' shaped $\phi_{\mathrm{SiO}}$ and Stokes $V$ profiles; reminiscent
of a typical Zeeman circular polarization spectrum. The trend is reversed
in the last two panels where the population at $3\;\mathrm{km\: s^{-1}}$
becomes increasingly more important in similar proportions, yielding
analogous but inverted profiles. We therefore find that both antisymmetric
``S'' and symmetric ``$\cup$'' or ``$\cap$'' shaped Stokes
$V$ profiles naturally arise from the model of \citet{Houde2013}
when anisotropic resonant scattering is allowed to occur externally
to the maser (in velocity space). 

Similar results are obtained for masers at the $\mathrm{SiO}\;\left(v=1,\: J=1\rightarrow0\right)$
transition at 43.1 GHz.

\section{Discussion\label{sec:Discussion}}

\begin{figure}
\epsscale{0.8}\plotone{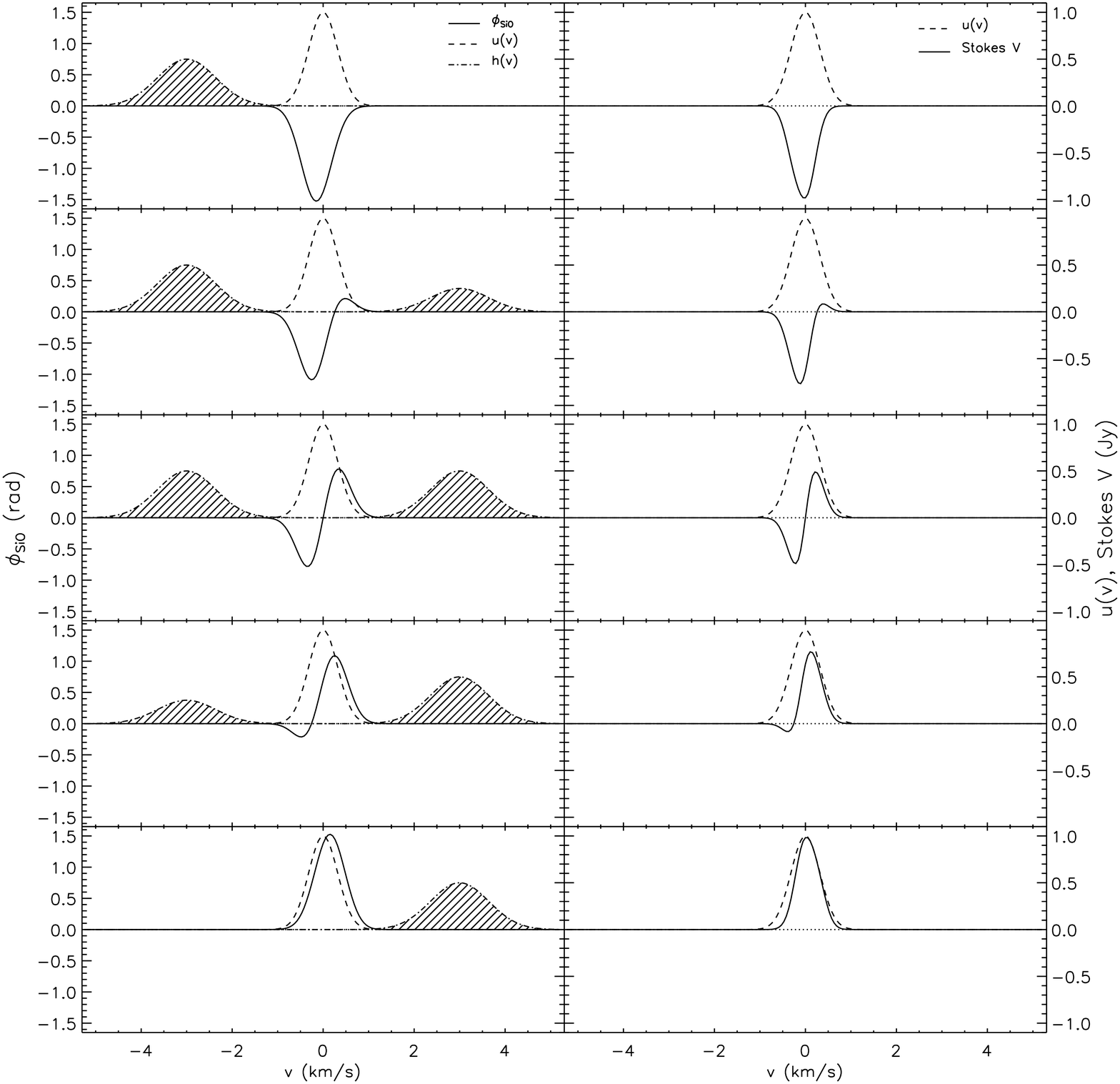}

\protect\caption{\label{fig:SiO_cp}Circular polarization spectra of the $\mathrm{SiO}\;\left(v=2,\: J=1\rightarrow0\right)$
transition at 42.8 GHz using the anisotropic resonant scattering model
of \citet{Houde2013}. The left panels show the relative phase shift
$\phi_{\mathrm{SiO}}$ between the linear polarization components
parallel and perpendicular to the orientation of the magnetic field
in the plane of the sky (solid curve; using the vertical scale on
the left), the incident linear polarization energy density ($u\left(v\right)$,
broken curve; using the vertical scale on the right), and the population
of scattering molecules ($h\left(v\right)$, dotted-broken and hashed
curve; normalized for display purposes). These were calculated using
the model of \citet{Decin2010} for the CSE of IK Tau. For the background
gas where the maser radiation originates (in the inner CSE) we set
$n_{\mathrm{H}_{2}}=10^{10}\;\mathrm{cm}^{-3}$, $T_{\mathrm{kin}}=1000\;\mathrm{K}$,
while in the foreground where the resonant scattering takes place
(5 AU further in the CSE) we have $n_{\mathrm{H}_{2}}=7.5\times10^{8}\;\mathrm{cm}^{-3}$,
$T_{\mathrm{ex}}=700\;\mathrm{K}$, and \textbf{$B=15\;\mathrm{mG}$}.
In both regions the SiO abundance relative to $\mathrm{H}_{2}$ was
set to $1.6\times10^{-5}$. The foreground SiO scattering populations
therefore have a maximum density $n_{\mathrm{SiO}}=1.2\times10^{3}$
cm$^{-3}$ for any of the two velocity ranges (i.e., centered at $\pm3\;\mathrm{km\: s^{-1}}$).
The right panels show the resulting Stokes $V$ profile (solid curve)
and, once more, the incident linear polarization energy density ($u\left(v\right)$,
 broken curve); both curves use the vertical scale on the right. An
interaction length $l_{\mathrm{s}}=5\;\mathrm{AU}$  were used for
these calculations.}
\end{figure}

The results of our numerical calculations shown in Figure \ref{fig:SiO_cp}
establish that the anisotropic resonant scattering model of \citet{Houde2013}
provides a viable explanation for the IK Tau SiO $v=1$ and $v=2$,
$\left(J=1\rightarrow0\right)$ masers observations of \citet{Cotton2011}.
The key to reproducing the observed Stokes $V$ profiles is the realization
that resonant scattering can occur from interaction of the maser radiation
with populations of foreground molecules located outside the velocity
range covered by the maser. As is clearly demonstrated in Figure \ref{fig:SiO_cp},
depending on the arrangement of the relative populations in either
sides of the maser line  a sequence of symmetric ``$\cup$'' or
``$\cap$'' and antisymmetric ``S'' shaped spectral profiles naturally
arises. These Stokes $V$ spectra share the same characteristics as
those seen in \citet{Cotton2011} (see their Figure 5, for example).
For IK Tau the overall population of SiO molecules emits at velocities
covering from approximately $26\;\mathrm{km\: s}^{-1}$ to $40\;\mathrm{km\: s}^{-1}$
(see Figure 2 in \citealt{Cotton2011}), with narrow maser features
interspersed over that range. The possibility for maser radiation
to interact with molecules located at neighboring velocities is therefore
non-negligible. This is also consistent with the mapping of 22 GHz
water masers in this source (and others) by \citet{Richards2012},
where over-dense regions of gas containing a significant fraction
of the mass loss are seen to be located over a large range of distances
from the center of expansion. These clumps have sizes and velocities
that scale with their positions in the CSE and, although they were
mapped with water masers, are likely to also contain SiO gas. For
IK Tau in particular these clouds have radii measured to be on the
order of 1 to 2 AU at a distance of a few AU from the expansion center.\textbf{
}We also note that \citet{Leal-Ferreira2013} have successfully detected
circular polarization signals in such over-dense gas regions in IK
Tau (and other sources) using the 22 GHz water maser line. The observed
Stokes $V$ spectra are consistent with a simple Zeeman model (i.e.,
they display an anti-symmetric ``S'' profile) and yield magnetic
field strengths on the order of 100 mG, when thus interpreted. 

Although there are significant uncertainties on the parameters used
in our calculations (e.g., temperatures, densities, and the volume
of the region of interaction $\mathcal{V}_{\mathrm{int}}=l_{\gamma}^{2}l_{\mathrm{s}}$),
the model is relatively robust in reproducing the results as there
exists a significant domain of combinations that can lead to similar
conclusions. For example, the results shown in Figure \ref{fig:SiO_cp}
obtained with the following parameters in the foreground where the
polarization conversion takes place: $n_{\mathrm{H}_{2}}=7.5\times10^{8}\;\mathrm{cm}^{-3}$,
$T_{\mathrm{ex}}=700\;\mathrm{K}$ $B=15\;\mathrm{mG}$, and $l_{\mathrm{s}}=5\;\mathrm{AU}$
apply equally well for $n_{\mathrm{H}_{2}}=7.5\times10^{8}\;\mathrm{cm}^{-3}$,
$B\simeq50\;\mathrm{mG}$, and $l_{\mathrm{s}}=0.5\;\mathrm{AU}$
or $n_{\mathrm{H}_{2}}=7.5\times10^{7}\;\mathrm{cm}^{-3}$, $B\simeq50\;\mathrm{mG}$,
and $l_{\mathrm{s}}=5\;\mathrm{AU}$, etc. Likewise, displacing the
populations of foreground molecules from $\pm3\:\mathrm{km\, s}^{-1}$
to $\pm4\:\mathrm{km\, s}^{-1}$ only requires increasing the magnetic
field strength from 15 mG to 30 mG to preserve the intensity and shape
of the Stokes $V$ spectra. Furthermore, the exact shape of the scattering
population distributions is largely irrelevant, only their numbers
(i.e., densities) matter. It is interesting to compare these magnetic
field strengths with those needed for the simple Zeeman model (i.e.,
1.9 to 6 G for IK Tau \citep{Herpin2006} or more in some cases \citep{Watson2009})
to match the measured levels of circular polarization (up to a few
tens of percent; see Figure 12 in \citealt{Cotton2011}). The amount
of circular polarization predicted by the scattering model is only
limited by that of the incident linear polarization; it is clear from
our results of Figure \ref{fig:SiO_cp} and equation (\ref{eq:phi})
that a minimal increase in magnetic field strength or adjustment in
other parameters could lead to a complete conversion of the linear
polarization. Although it is to be noted that gradients in the velocity
of the gas and magnetic field strength along the line of sight can
account for deviations from the usual antisymmetric ``S'' shape
of Stokes $V$ Zeeman spectra (for Zeeman sensitive species, especially,
this can even lead to the ``disappearance'' of spectral components
\citep{Watson2009}), the fact that the simple Zeeman model cannot
reproduce the symmetric Stokes $V$ profiles observed by\textbf{ }\citet{Cotton2011}\textbf{
}renders our model an even more viable alternative for explaining
the polarization characteristics of SiO masers in evolved stars. This
is because, as explained by \citet{Cotton2011}, the lack of significant
velocity structure for neighboring lines of sight in relation to the
overall range of velocities where masers were detected makes it unlikely
that such effects can explain their observations. In a more general
context we should also note that based on the strength of magnetic
fields predicted and observed alone, a hybrid of Zeeman and scattering
non-Zeeman process to explain the observed circular polarization spectra
should not be ruled out. For example, the field strength of a few
times 15 mG at 10 AU in the CSE of IK Tau needed for the resonant
scattering model could imply the presence of a surface magnetic field
of the order of 1 G when we adopt a $1/r^{2}$ functional form for
the magnetic field strength as a function of position in the CSE.
This would not necessarily be inconsistent with other results from
maser measurements. It is also interesting to note that a line-of-sight
surface magnetic field strength of $2-3\mathrm{\; G}$ has recently
been estimated from circular polarization measurements for another
Mira star (i.e., $\chi\mathrm{-Cigni}$) by \citet{Lebre2014}. Although
this value is consistent with our previous estimate using our calculations
for the SiO lines in IK Tau, \citet{Lebre2014} concluded, however,
that their results were more consistent with a $1/r$ than a $1/r^{2}$
magnetic field strength scaling.

On the other hand it is to be acknowledge that the models presented
here are probably too simple to fully account for the complexity of
the physical conditions found in the CSE of IK Tau (or other objects
in general), as well as for the range of existing observations. One
notable example concerning the observations of \citet{Cotton2011}
is the case of Regions B and C for their SiO $v=2$, $\left(J=1\rightarrow0\right)$
42.8 GHz observations. Considering that the two masers occur at the
same velocity (i.e., $\simeq39\;\mathrm{km\: s^{-1}}$) and that they
are spatially separated by approximately 1 AU (as projected on the
sky; see Fig. 4 of \citealt{Cotton2011}), the apparent similarity
in their environments may make it difficult to reconcile the fact
that their Stokes $V$ spectra are inverted from one another (see
Fig. 5 of \citealt{Cotton2011}). That is, the Stokes $V$ spectrum
of Region B is similar to the top spectrum in our Figure \ref{fig:SiO_cp},
while that of Region C resembles the bottom spectrum. We can put forth
a simple solution to this problem by noting that an inversion of a
given Stokes $V$ spectrum does not have to result from a displacement
of the scattering foreground molecules from, say, the low side to
the high side of the incident radiation in velocity space, but can
also result by a change in the relative alignment between the linear
polarization state of the incident radiation and the foreground magnetic
field. Since the overall sign of the spectrum is controlled through
the $2\alpha_{0}\beta_{0}=\sin\left(2\theta\right)$ factor in equation
(\ref{eq:v}), the needed angular rotation for the sign change can
be small if $\theta$ is small to start with, or at most as large
as $\pi/4$ if $\theta$ is initially close to $\pi/4$. Furthermore,
these potential changes in $\theta$ may also be required in a more
general sense since it is unlikely that SiO maser radiation blue-shifted
relative to the velocity center can interact with both relatively
blue- and red-shifted scattering gas in the CSE, as implied in our
Figure \ref{fig:SiO_cp}. That is, since the scattering gas must be
located between the maser formation region and the observer the nature
of the expanding wind in the CSE implies that blue-shifted maser radiation
reaching the observer can only resonantly scatter with gas that is
further blue-shifted in velocity space. Evidently the presence of
turbulent, infall, or rotational velocity components may alter this
picture. Nonetheless, changes in the relative orientation between
the foreground magnetic field and the linear polarization state of
the incident maser radiation for different pockets of scattering gas
along the line of sight may therefore be needed to account for all
observations. For red-shifted maser emission, it is also possible
that blue- and red-shifted scattering gas would not be subjected to
the same physical conditions in view of their respective locations
in the CSE. For simplicity, we have not accounted for these differences
in our calculations. In all cases the added complexity needed in our
model resides in accurately knowing a set of physical parameters to
a level of precision we do not possess. 

Incidentally, the fact that red-shifted masers can scatter with both
relatively blue- and red-shifted gas and have to propagate through
a larger column of gas in the CSE before reaching the observer in
comparison to blue-shifted masers (i.e., they are located further
away along the line of sight) would lead us to expect that the presence
of circular polarization should be more prevalent in the spectra of
red-shifted masers. It is therefore interesting to note that most
of the spectra for the two transitions presented in \citet{Cotton2011}
are red-shifted (i.e., five out of six for each transition, only Regions
F are blue-shifted).

We also note that a previous observational comparison of circular
polarization levels between the 43.1 GHz $\left(v=1,\; J=1\rightarrow0\right)$
and 86.2 GHz $\left(v=1,\; J=2\rightarrow1\right)$ SiO maser lines
for VY CMa \citep{McIntosh1994} revealed that a corresponding ratio
of the Stokes $V$ intensities between the two spectra of $\sim2$,
consistent with the Zeeman effect. A similar argument can be made
through a comparison of the results of \citet{Herpin2006} for IK
Tau for the 86.2 GHz $\left(v=1,\; J=2\rightarrow1\right)$ line with
those of \citet{Cotton2011} at 43.1 GHz $\left(v=1,\; J=1\rightarrow0\right)$.
That is, \citet{Herpin2006} find the average ratio of circular to
linear polarization to be $\sim0.3$ at the higher frequency, while
\citet{Cotton2011} measure it to be $0.13$ at 43.1 GHz. We find
again a ratio neighboring $2$ favoring the lower frequency transition.
We performed similar calculations as those shown in Figure \ref{fig:SiO_cp}
for these two $v=1$ transitions, as well as for the 85.6 GHz $\left(v=2,\; J=2\rightarrow1\right)$
line since our previous calculations of Figure \ref{fig:SiO_cp} were
made for the 42.8 GHz $\left(v=2,\; J=1\rightarrow0\right)$ line,
and compared the Stokes $V$ intensities. The ratio of the $\left(J=1\rightarrow0\right)$
to $\left(J=2\rightarrow1\right)$ intensities varies somewhat depending
on the case chosen (i.e., what panel of Fig. \ref{fig:SiO_cp} is
used) but we find that it is also close to 2 for both pairs of lines.
For example, the 42.8-GHz/85.6-GHz ratio for the top panel of Figure
\ref{fig:SiO_cp} the peak Stokes $V$ intensity numbers are $\simeq-1:-0.6$,
while for the middle panel they are $\simeq0.49:0.23$. It thus appears
that the ratios obtained for the resonant scattering model are also
consistent with the aforementioned observations.

\begin{figure}
\epsscale{1.0}\plotone{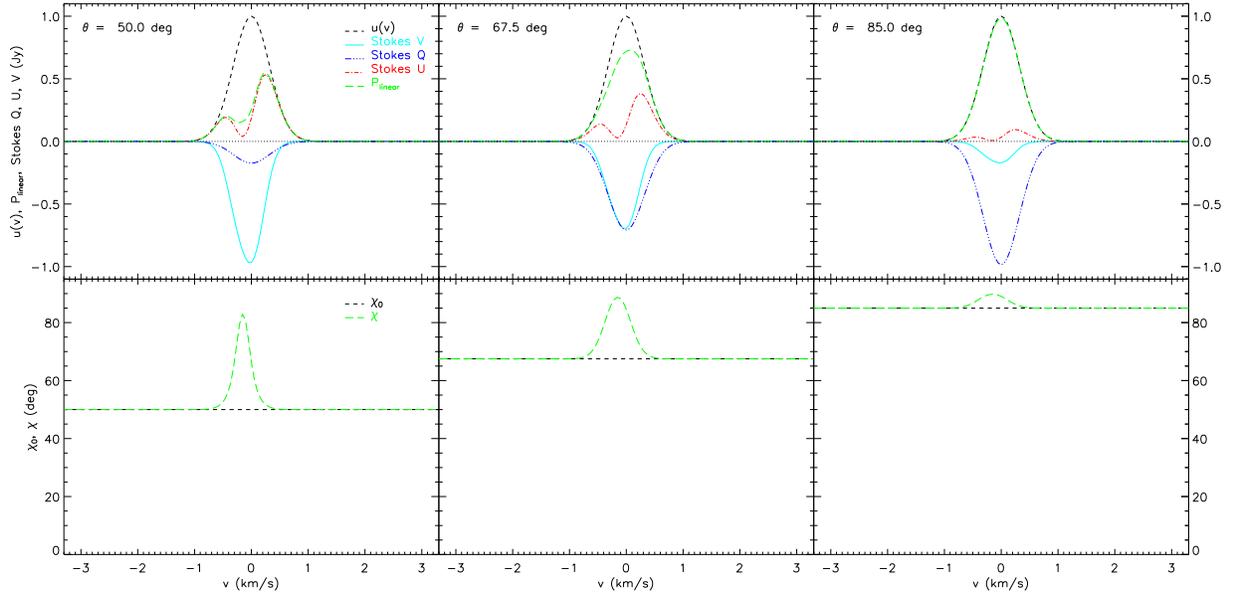}

\protect\caption{\label{fig:stokes-1}Stokes $V,$ $Q$, $U$ parameters, and linear
polarization signals (top) and angles (bottom) for the scattering
distribution shown in the top panel of Figure \ref{fig:SiO_cp} for,
from left to right, $\theta=50\:\deg$, $67.5\:\deg$, and $85\:\deg$.
The incident linear polarization maser signal $u\left(v\right)$ is
oriented at an angle $\theta\left(=\chi_{0}\right)$ relative to the
foreground magnetic field. After resonantly scattering with the foreground
gas the Stokes $V,$ $Q$, and $U$ radiation signals emerge. The
linear polarization signal is then $P_{\mathrm{linear}}$ oriented
at an angle $\chi$ relative to the foreground magnetic field (positive
Stokes $Q$ is aligned with the magnetic field). }
\end{figure}

\begin{figure}
\epsscale{1.0}\plotone{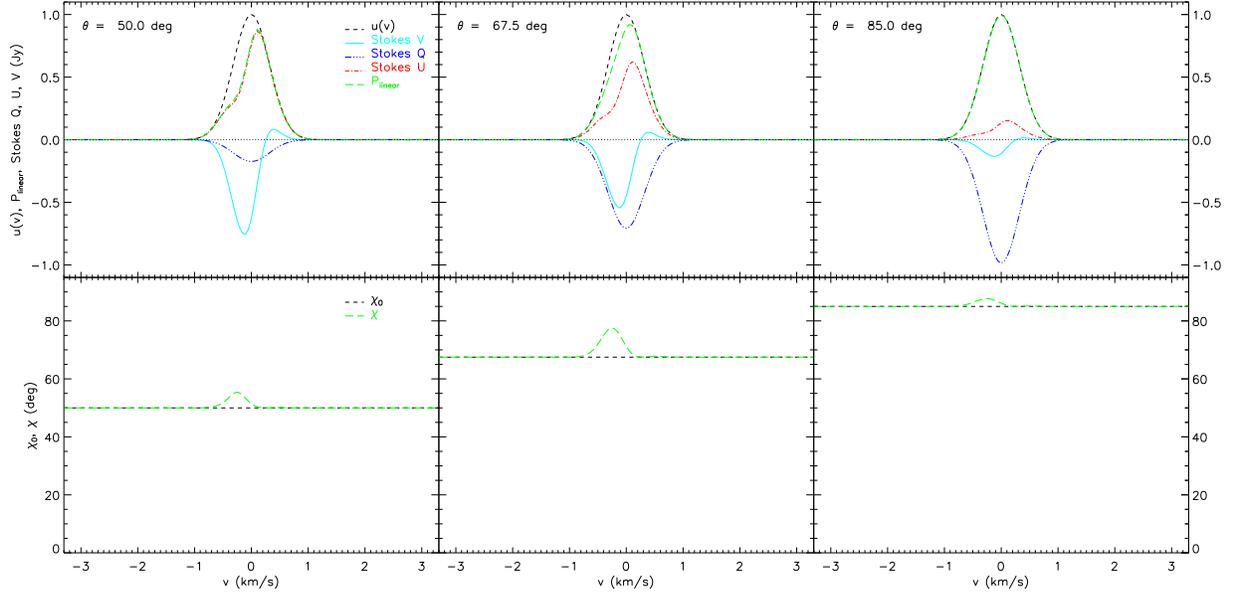}

\protect\caption{\label{fig:stokes-2}Same as figure \ref{fig:stokes-1} but for case
corresponding to the second panel from the top of Figure \ref{fig:SiO_cp}. }
\end{figure}

\begin{figure}
\epsscale{1.0}\plotone{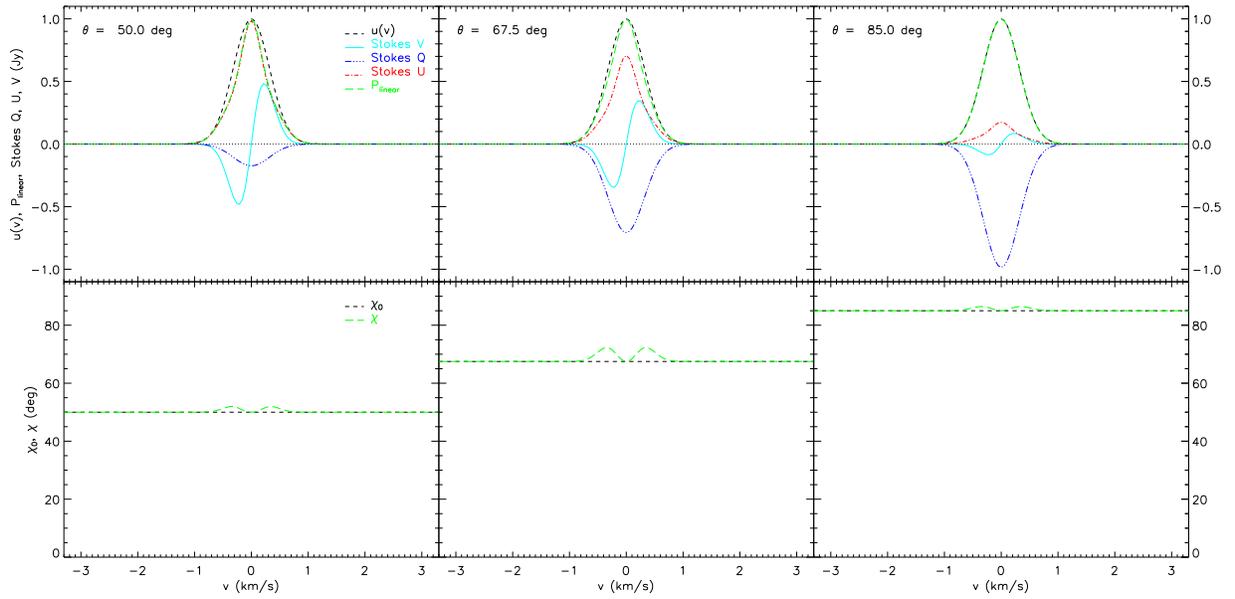}

\protect\caption{\label{fig:stokes-3}Same as figure \ref{fig:stokes-1} but for case
corresponding to the middle panel of Figure \ref{fig:SiO_cp}. }
\end{figure}

A further test that can be applied to our model consists of producing
corresponding linear polarization spectra to the results shown in
Figure \ref{fig:SiO_cp}. This was done using equations (\ref{eq:q})-(\ref{eq:chi})
and the resulting spectra are shown in Figures \ref{fig:stokes-1}
to \ref{fig:stokes-3}. Care must be taken when comparing these spectra
with the observations of \citet{Cotton2011} since their spectra are
those obtained after resonant scattering has taken place. The amount
of rotation induced on the linear polarization signal is a function
of the amount of scattering taking place (in the phase shift $\phi$
of eq. (\ref{eq:chi-chi0})), as well as the initial orientation of
the incident linear polarization in relation to the foreground magnetic
field before scattering (i.e., the angle $\theta$). Because we do
not a priori know these two orientations it is impossible to determine
the precise conditions that led to the spectra presented in \citet{Cotton2011}.
Still, it may be possible to test, under simple assumptions, whether
some spectral features predicted by our model are realized observationally.

More precisely, for the results presented in Figures \ref{fig:stokes-1}
to \ref{fig:stokes-3} we have produced the Stokes $V$, $Q$, and
$U$ spectra, as well as calculated the linear polarization $P_{\mathrm{linear}}=\left(Q^{2}+U^{2}\right)^{1/2}$
and angles $\chi=0.5\arctan\left(U/Q\right)$ for different relative
orientations between the incident linear polarization (of amplitude
$u\left(v\right)$ and angle $\chi_{0}$ in the figures) and the foreground
magnetic field. These orientations were set at $\theta=50$, $67.5$,
and $85\;\mathrm{deg}$. At $\theta=45\;\mathrm{deg}$ all of the
incident linear polarization is in Stokes $U_{0}$ and the conversion
effect will most strongly affect the shape of $P_{\mathrm{linear}}$,
however the angle $\chi$ will not show any features since Stokes
$Q$ is then zero; choosing a slightly displaced orientation at $\theta=50\;\mathrm{deg}$
allows us to better observe the effect in that neighborhood. Conversely,
the conversion effect is inexistent at $\theta=90\;\mathrm{deg}$
since then all of the incident linear polarization is in Stokes $Q_{0}$;
we therefore chose $\theta=85\;\mathrm{deg}$. Most importantly, we
emphasize that \emph{the angle $\theta$ was assumed constant across
the spectral line}. Evidently, this not need to be the case in reality,
and variations in $\theta$ as a function of velocity would alter
the spectral shapes produced in our calculations. 

Figure \ref{fig:stokes-1} corresponds to the top panel of Figure
\ref{fig:SiO_cp} (symmetric Stokes $V$), while Figures \ref{fig:stokes-2}
and \ref{fig:stokes-3} are for the second (from the top) and middle
panels (anti-symmetric Stokes $V$). One can see that significant
deformations in the linear polarization spectra $P_{\mathrm{linear}}$
can be observed when most of the incident linear polarization is in
Stokes $U$ for the cases of symmetric Stokes $V$ profiles (see $\theta=50\;\mathrm{deg}$
in Fig. \ref{fig:stokes-1}). In that case, a ``blue'' shoulder
appears in the linear polarization spectrum. We note that some of
\citet{Cotton2011} spectra appear to share some of that characteristic
(e.g., Region D in their Fig. 5). On the other hand, the polarization
angle $\chi$ for that same case shows important variations over a
larger range of orientations $\theta$ (e.g., several tens of degrees
when $\theta=50$ and $\theta=67.5\;\mathrm{deg}$ in Fig. \ref{fig:stokes-1}).
Again, some of the spectra of \citet{Cotton2011} also display large
variations in polarization angles (sometimes more severe), although
it is difficult to pick a specific pattern from their data. The same
types of behavior are observed in Figure \ref{fig:stokes-2} but to
a lesser degree.

The anti-symmetric Stokes $V$ case of Figure \ref{fig:stokes-3}
shows that the shape of the linear polarization spectrum $P_{\mathrm{linear}}$
is not significantly impacted by the presence of circular polarization
for the set of parameters used in our calculations. Interestingly,
the corresponding pattern for the polarization angle $\chi$ is similar
in shape to the one found for Region A in Figure 5 of \citet{Cotton2011}.
This is more easily seen through a comparison with the middle panel
of Figure \ref{fig:stokes-3} when $\theta=67.5\;\mathrm{deg}$. The
amplitude of the variation could to some extent be adjusted by tuning
the parameters (e.g., slightly changing the strength of the magnetic
field). Also, the range of values for $\chi$ could be shifted up
or down by a simple rotation of the reference frame used to calculate
Stokes $Q$ and $U$ with respect to the orientation of the foreground
magnetic field.

It remains to generalize and apply the anisotropic resonant scattering
model to other maser transitions such as for water and methanol, which
unlike SiO are asymmetric top molecules. It will be interesting to
find out if, for example, the polarization characteristics of the
6.7-GHz methanol maser can be satisfactorily reproduced. Given the
current uncertainty concerning the Zeeman sensitivity of that transition
\citep{Vlemmings2011}, physical models requiring weaker magnetic
field strengths may become necessary.

\section{Conclusion\label{sec:Conclusion}}

In this paper we applied the anisotropic resonant scattering model
developed by \citet{Houde2013} to explain the presence of non-Zeeman
circular polarization signals recently detected in $^{12}\mathrm{CO}\;\left(J=2\rightarrow1\right)$
and $\left(J=1\rightarrow0\right)$ transitions in molecular clouds
\citep{Houde2013,Hezareh2013} to Stokes $V$ spectra of SiO $v=1$
and $v=2$, $\left(J=1\rightarrow0\right)$ masers commonly observed
in evolved stars. It was found that the observed antisymmetric ``S''
and symmetric ``$\cup$'' or ``$\cap$'' shaped spectral profiles
naturally arise when the maser radiation scatters off populations
of foreground molecules located outside the velocity range covered
by the maser. Using typical values for the relevant physical parameters,
it is estimated that magnetic field strengths on the order of a few
times 15 mG are sufficient to explain the observational results previously
published by \citet{Cotton2011} for the evolved star IK Tau.

\acknowledgements{We are grateful to T. Hezareh and H. Wiesemeyer for helpful discussions.
M.H.'s research is funded through the NSERC Discovery Grant, Canada
Research Chair, and Western's Academic Development Fund programs. }

\end{document}